\pdfoutput=1
\newcommand*{\ATLASLATEXPATH}{}
\documentclass[cernpreprint, texlive=2016, USenglish]{\ATLASLATEXPATH atlasdoc}
\usepackage{\ATLASLATEXPATH atlaspackage}
\usepackage{\ATLASLATEXPATH atlasbiblatex}
\usepackage{\ATLASLATEXPATH atlasphysics}
\DeclareUnicodeCharacter{2212}{-}

\AtlasTitle{Search for magnetic monopoles and stable high-electric-charge objects in 13 TeV proton--proton collisions with the ATLAS detector}

\AtlasAbstract{%
A search for magnetic monopoles and high-electric-charge objects is
presented using 34.4~fb$^{-1}$ of 13~TeV $pp$ collision data collected
by the ATLAS detector at the LHC during 2015 and 2016. The considered
signature is based upon high ionization in the transition radiation
tracker of the inner detector associated with a pencil-shape energy
deposit in the electromagnetic calorimeter. The data were collected by
a dedicated trigger based on the tracker high-threshold hit
capability. The results are interpreted in models of Drell--Yan pair
production of stable particles with two spin hypotheses (0 and 1/2)
and masses ranging from 200~GeV to 4000~GeV. The search improves by
approximately a factor of five the constraints on the direct
production of magnetic monopoles carrying one or two Dirac magnetic
charges and stable objects with electric charge in the range
$20\le|z|\le60$ and extends the charge range to $60<|z|\le100$.  }

\AtlasRefCode{EXOT-2017-20}

\PreprintIdNumber{CERN-EP-2019-084}

\arXivId{1905.10130}

\HepDataRecord{10.17182/hepdata.89874}

\AtlasJournalRef{Phys.\ Rev.\ Lett. 124 (2020) 031802}
                                   
\AtlasDOI{10.1103/PhysRevLett.124.031802}

 \AtlasCoverSupportingNote{Monopole/HECO search 13 TeV 2016}{https://cds.cern.ch/record/2267571}

\hypersetup{pdftitle={ATLAS document},pdfauthor={The ATLAS Collaboration}}

\begin{document}

\maketitle

The symmetry between electric and magnetic charge in Maxwell's
Equations and the explanation for electric charge quantization
resulting from Dirac's quantum description of the magnetic
monopole~\cite{Dirac1931,Dirac1948} are compelling arguments for its
existence. Neither the spin nor the mass of a Dirac monopole is
theoretically constrained. While monopoles appearing in grand
unification theories~\cite{tHooft1974,Polyakov1974} typically have
masses of the order of the unification scale ($m\sim 10^{16}$~GeV),
some extensions of the Standard Model predict \textit{electroweak}
monopoles with masses as low as
4~TeV~\cite{Cho1997,Cho2015,Ellis2016,Arunasalam2017,Ellis2017}.
TeV-mass monopoles can be produced in the early Universe thermally or via
the Kibble mechanism~\cite{Kibble1976,Preskill1979} in cosmological
scenarios with a low reheat temperature after
inflation~\cite{GouldRajantie2017}.

Dirac's argument predicts the
fundamental magnetic charge to be $q_\text{m} =
Ng_\text{D}ec$,~\footnote{In this definition, $q_\text{m}$ is in SI
units and $g_\text{D}$ is a dimensionless quantity.} where $g_\text{D}
= 1/(2\alpha) = 68.5$ is the Dirac charge, $\alpha$ is the fine
structure constant, $N$ is an integer number, $e$ is the unsigned
electron charge and $c$ is the speed of light in vacuum. This implies
that a high-velocity Dirac monopole of magnetic charge
$|g|=g_\text{D}$ would interact with matter in a manner similar to
that of an ion of electric charge $|z|=68.5$, where $z$ is in units of
$e$. Since the energy loss is proportional to the square of the
charge, a monopole with $|g|=g_\text{D}$ would deposit 4700 times more
energy by ionization than a proton. The high stopping power also
results in the production of a large number of $\delta$-rays. These
two features result in a high-ionization signature that is also
expected in the case of exotic stable high-electric-charge objects
(HECOs), which may include, for example, aggregates of
$ud$-~\cite{Holdom2018} or $s$-quark matter~\cite{Farhi1984},
$Q$-balls~\cite{Coleman1985,Kusenko1998}, and micro black-hole
remnants~\cite{Koch2007}.

This Letter presents a search for magnetic monopoles and HECOs,
collectively referred to as highly ionizing particles, or HIPs, using
34.4~fb$^{-1}$ of 13~TeV proton--proton ($pp$) collision data
collected by the ATLAS detector at the CERN Large Hadron Collider
(LHC) during 2015 and 2016. Events containing at least one
high-ionization object are selected. The results are interpreted in
models of spin-0 and spin-1/2 Drell--Yan pair production of stable
particles carrying one or two Dirac magnetic charges or an electric
charge in the range $20\le|z|\le100$ with masses ranging from 200~GeV
to 4000~GeV.

Should monopoles exist in the mass range accessible to a particle
accelerator, they could be copiously produced at the LHC. If they were
detected, the measured mass and coupling would severely restrict
cosmological scenarios. Since the numerous searches for monopoles of
cosmological origin in cosmic rays and in
matter~\cite{Burdin2015,IceCube2015} have limited sensitivity to
TeV-mass HIPs, the cross-section limits for low-mass HIPs from
searches at
colliders~\cite{Fairbairn2007,CDF2006,STAR2007,OPAL2008,EXOT-2010-04,EXOT-2011-27,EXOT-2012-10,EXOT-2014-16,MoEDAL2016a,MoEDAL2017a,MoEDAL2017b,MoEDAL2019,EXOT-2013-20,CMS2013a,EXOT-2017-13}
are 6--9 orders of magnitude more stringent. The first LHC searches
for HECOs and monopoles were made by the ATLAS Collaboration in 8~TeV
$pp$ collisions~\cite{EXOT-2010-04,EXOT-2011-27} by exploiting the
high-ionization signature. The higher collision energy, the five times
larger dataset and improvements in the trigger extend the sensitivity
of the present search. While the previous search studied
$|g|\le1.5g_\text{D}$, the present search considers monopoles up to
$|g|=2g_\text{D}$, which are motivated by Schwinger, who showed that
$N$ must be even for particles possessing both electric and magnetic
charge~\cite{Schwinger1966,Schwinger1968,Schwinger1969}. The ATLAS
monopole searches~\cite{EXOT-2011-27,EXOT-2014-16} are complementary
to those performed using the dedicated MoEDAL
experiment~\cite{MoEDAL2016a,MoEDAL2017a,MoEDAL2017b,MoEDAL2019},
which uses an induction technique to detect the magnetic flux of
monopoles trapped in matter. While MoEDAL is sensitive to magnetic
charges up to $5g_\text{D}$, the present ATLAS search is able to set
significantly better cross-section constraints for $1g_\text{D}$ and
$2g_\text{D}$, the charge range in which it has a good
acceptance~\cite{DeRoeck2012a}.

Unlike searches using the induction technique, the present search is
sensitive to high-charge HECOs in addition to monopoles. It is
complementary to the low-charge HECO searches performed by ATLAS
($2\leq|z|\leq 7$~\cite{EXOT-2012-10,EXOT-2013-20,EXOT-2017-13}) and
CMS ($2\leq|z|\leq 8$~\cite{CMS2013a}), all of which used muon
triggers. A muon trigger is not appropriate for high-charge HECOs,
which typically stop in the electromagnetic calorimeter, due to the
charge-square dependence of $dE/dx$. In LHC Run~1, ATLAS probed
electric charges up to $|z|=60$~\cite{EXOT-2010-04,EXOT-2014-16} via
the high-ionization signature. The present analysis is able to probe
HECOs up to $|z|=100$, thereby reaching the previously unexplored
charge range predicted for $ud$-quark matter~\cite{Holdom2018}.

The present search exploits the very characteristic high-ionization
signature of HIPs in the ATLAS detector~\cite{PERF-2007-01}. The ATLAS
transition radiation tracker (TRT), which is the outermost tracker of
the inner detector, consists of a barrel ($|\eta|<1.0$ and radius
0.563~m~$<r<1.066$~m)~\footnote{ATLAS uses a right-handed coordinate
system with its origin at the nominal interaction point (IP) in the
center of the detector and the $z$-axis coinciding with the axis of
the beam pipe. The $x$-axis points from the IP to the center of the
LHC ring, and the $y$-axis points upward. Cylindrical coordinates
($r$, $\phi$) are used in the transverse plane, $\phi$ being the
azimuthal angle around the beam pipe. The pseudorapidity is defined in
terms of the polar angle $\theta$ as $\eta=-\ln \tan(\theta/2)$.} with
4-mm-diameter straws oriented parallel to the beam-line, and two
endcaps ($0.77<|\eta|<2.0$) with straws oriented radially.  In LHC
Run~2, 56\% of the straws were filled with xenon gas while the others
were filled with argon gas.  Energy deposits in a TRT straw greater
than 200~eV, called \textit{low-threshold} (LT) hits, are used for
tracking. The \textit{high-threshold} (HT) hits, which result from
energy deposits exceeding 6~keV in Xe (2~keV in Ar) are typically used
for electron identification, but can also indicate the presence of a
highly ionizing particle. A 2~T superconducting solenoid surrounds the
TRT. The lead/liquid-Ar (LAr) barrel electromagnetic (EM) calorimeter
lies outside the solenoid in the $|\eta|<1.475$ region. It is divided
into three shower-depth layers: EM1, EM2 and EM3, with accordion-shape
electrodes and lead absorbers. The EM2 layer has the largest sampling
depth ($16$ to $20X_0$); its cell granularity is
$\Delta \eta \times \Delta \phi = 0.025\times0.025$. In
the \mbox{$|\eta|<1.8$} region, an additional presampler layer is used
to measure the energy lost in front of the calorimeter.  As a HIP
traverses the TRT, a localized region of high ionization density with
many HT hits from both the HIP and the $\delta$-rays is produced in
its wake. HIPs slow down (and usually stop) in the EM calorimeter,
where they leave a pencil-shape energy deposit since they do not
induce a shower, being much heavier than the electron.

Signal efficiency estimates rely heavily on simulations,
which use the \textsc{Geant}4 framework~\cite{Geant42003,SOFT-2010-01}
to model the HIP propagation and behavior in ATLAS. The detector
simulation includes the full ATLAS geometry, descriptions of monopole
acceleration in the solenoidal magnetic field, ionization energy
losses in matter~\cite{Ahlen1978,Ahlen1980,Ahlen1982,Cecchini2016},
$\delta$-electron production along the HIP trajectory, and a model
accounting for electron--ion recombination in the LAr EM
calorimeter~\cite{Burdin2011}. For monopoles, the trajectory in the
solenoidal magnetic field is straight in the $r$--$\phi$ plane and
bends in the $r$--$z$ plane and the Bethe--Bloch formula is
modified~\cite{Ahlen1978,Ahlen1980} to account for the
velocity-dependent Lorentz force. The interaction of HIPs with matter
is independent of HIP spin.

The Drell--Yan (DY) pair-production process is used to estimate the
kinematic distributions and cross sections of spin-0 and spin-1/2 HIPs
in the relevant ranges of charge and mass.
The \textsc{MadGraph}5\_aMC@NLO~\cite{Alwall2014} event generator was
used to model leading-order HIP DY pair production from the initial
$pp$ state via quark--antiquark annihilation into a virtual photon.
The charge-squared dependence of the HIP
coupling to the photon implies divergences in the
perturbative expansion beyond leading
order~\cite{Milton2006}. \textsc{Pythia} version
8.212~\cite{Pythia6,Pythia8} was employed for the hadronization and
the underlying-event generation, using the NNPDF2.3~\cite{NNPDF23LO}
parton distribution functions of the proton with the
A14~\cite{ATL-PHYS-PUB-2014-021} set of tuned parameters (`tune'). The
A2~\cite{ATL-PHYS-PUB-2012-003} tune was used for the `pileup', which are
additional simulated $pp$ collisions overlaid on each event according
to the distribution of the number of $pp$ interactions per bunch
crossing, $\mu$, in the data.  Fully simulated HIP Monte Carlo (MC)
samples are computationally intensive due to the high ionization. To
minimize the number of such MC samples, model-independent efficiency
maps, finely binned in kinetic energy and $|\eta|$, are produced from
fully simulated single-particle samples of a given mass and
charge. The spin-0 DY HIP four-vectors from the generator are used to
sample the maps in order to derive the spin-0 DY HIP selection
efficiencies. The DY spin-1/2 selection efficiencies are derived from
fully simulated DY samples, which are also used to validate the
results obtained by sampling the efficiency maps and to assign a
modeling uncertainty. The selection efficiencies for spin-0 DY HIPs
are higher than for spin-1/2 because they have more central $\eta$ and
harder kinetic energy distributions.

Incomplete knowledge of the simulation parameters translates into
systematic uncertainties in the signal efficiencies. These
uncertainties are estimated by varying the parameters within a range
corresponding to an uncertainty of one standard deviation, as
described in more detail in Ref.~\cite{EXOT-2014-16}. The dominant
uncertainties are those due to the descriptions of the detector
material and of electron--ion recombination in the EM calorimeter,
each with an average relative uncertainty of 7\%. Other relevant
parameters include the dependence of the multiplicity of TRT LT hits
on pileup $\mu$, the HIP energy-loss calculation as well as the yield
and range of $\delta$-electrons, and the fraction of energy cross-talk
between adjacent EM calorimeter cells. Additional uncertainties
unrelated to detector effects include the uncertainty from MC
statistics and the systematic uncertainty due to imprecise modeling or
deriving DY efficiencies from the single-particle maps.

ATLAS uses a two-level trigger system~\cite{TRIG-2016-01}.
Level~1 is a hardware-based trigger that defines calorimeter `regions
of interest' (RoI). A dedicated software-based high-level trigger
(HLT), which imposes requirements on the number and fraction of TRT HT
hits in a narrow region around the RoI, started collecting data in
October 2015. The level-1 seed requires transverse energy
$E_\text{T}>22$~GeV in the electromagnetic calorimeter. RoI
candidates with $E_\text{T}<50$~GeV in conjunction with an
$E_\text{T}$-dependent minimum of 1 to 2~GeV of energy in the
hadronic calorimeter are vetoed. This unavoidable \textit{hadronic
veto} requirement, which improves rejection of hadrons in analyses
using electrons and photons, limits the sensitivity to low-charge
HIPs, which often penetrate to the hadronic calorimeter.  In contrast
to the 8~TeV search~\cite{EXOT-2014-16}, this veto on an energy deposit in
the hadronic calorimeter is not applied to RoI candidates with
$E_\text{T}>50$~GeV, thereby increasing the acceptance for the HIPs
that do not stop in the EM calorimeter. Following the level-1 trigger,
the HLT is used to select HIP candidates with HT hits in the TRT. The
counting is done in an $r$--$\phi$ wedge
of size 10~mrad, where two trigger variables, $N_\text{HT,trig}$ and
$f_\text{HT,trig}$, are used to define the number and fraction of
TRT HT hits, respectively. The HLT selection criteria are defined as
$N_\text{HT,trig} > 30$ and $f_\text{HT,trig} > 0.5$ to control the
rate while maintaining high signal efficiency for high-charge HIPs. In
addition, a pseudorapidity requirement of $|\eta|<1.7$ is applied to
avoid the forward regions, which contain more backgrounds due to
multijet events producing a higher rate of level-1 trigger seeds.

In this search, the HIP signal sensitivity is governed by the DY
kinematic distributions and the dependence of the stopping power on
two variables: the square of the charge and the velocity relative to
the speed-of-light $\beta$. For HECOs, $dE/dx$ is proportional to
$1/\beta^2$ whereas it varies as $\ln \beta^2$ for
monopoles~\cite{Ahlen1978,Ahlen1980}. The main source of signal loss
is due to HIPs failing to produce a level-1 trigger RoI, because they
either stop before the EM calorimeter (e.g., high-charge HIPs),
deposit too little energy in the calorimeter or penetrate to the
hadronic calorimeter and invoke the veto (e.g., low-charge HIPs). The
probability for a HIP from DY pair production to induce a level-1
trigger signal is around 60\% for HECOs with $|z|=20$ and 55\% for
monopoles with $|g|=g_\text{D}$ and decreases to 18\% for $|z|=60$ and
10\% for $|g|=2g_\text{D}$. For HIPs with $|\eta|<1.375$ that pass the
level-1 trigger, the efficiency to satisfy the HLT and the remaining
offline selection, described below, is generally 25\% to 60\%.

Any data that fail the electron--photon data quality requirements are
discarded as are events flagged as containing noise in the LAr
calorimeter. The event selection then starts by identifying events
containing at least one candidate featuring a topological cluster of
EM calorimeter cells~\cite{PERF-2014-07} with $E_\text{T}>18$~GeV in
the $|\eta| < 1.375$ region.  The remaining selection is based on two
powerful background-discriminating variables, denoted $f_\text{HT}$
and $w$.  The selected EM cluster candidates are used to seed the
$f_\text{HT}$ variable, which is similar to the trigger variable
$f_\text{HT,trig}$ except that an 8-mm-wide rectangular road is used
instead of a 10-mrad wedge, to better confine the hit counting to the
region closest to the HIP trajectory.  The $w$ variable gives a
measure of the lateral energy dispersion of the EM cluster
candidate. For each EM cluster candidate, the associated energy
contained in the presampler, EM1 layer and EM2 layer is denoted by
$E_{0}$, $E_{1}$ and $E_{2}$, respectively. Three $w_i$ variables
($i=0$, 1, 2) are defined as the fraction of EM cluster energy $E_i$
contained in the two most energetic cells in the presampler, the four
most energetic cells in EM1, and the five most energetic cells in EM2,
respectively. If the cluster energy in layer $E_{i}$ is confined to a
single cell, then $w_i=1$, consistent with the narrow shower
expected for HIPs. In each layer, the number of cells is chosen to
optimize the signal efficiency and the discrimination power between
HIPs and electron or jet backgrounds. The combined lateral energy
dispersion $w$ is thus defined as the average of all $w_i$ ($i=0$, 1,
2) for which $E_{i}$ exceeds 10~GeV (relaxed to 5~GeV for $i=2$). The
latter requirement ensures that only layers with energy deposits
significantly above the cluster-level noise, which depends on both the
cell-level noise and the cell granularity, contribute to the $w$
computation. In addition, at least one of the $E_{0}$ and $E_{1}$
requirements must be satisfied.  The final selection requirements are
$f_\text{HT}\geq 0.7$ and $w\geq 0.96$, a choice which maximizes the
signal-to-background ratio for the majority of the signal samples.

\begin{figure}[tb]
  \begin{center} \includegraphics[width=1\linewidth]{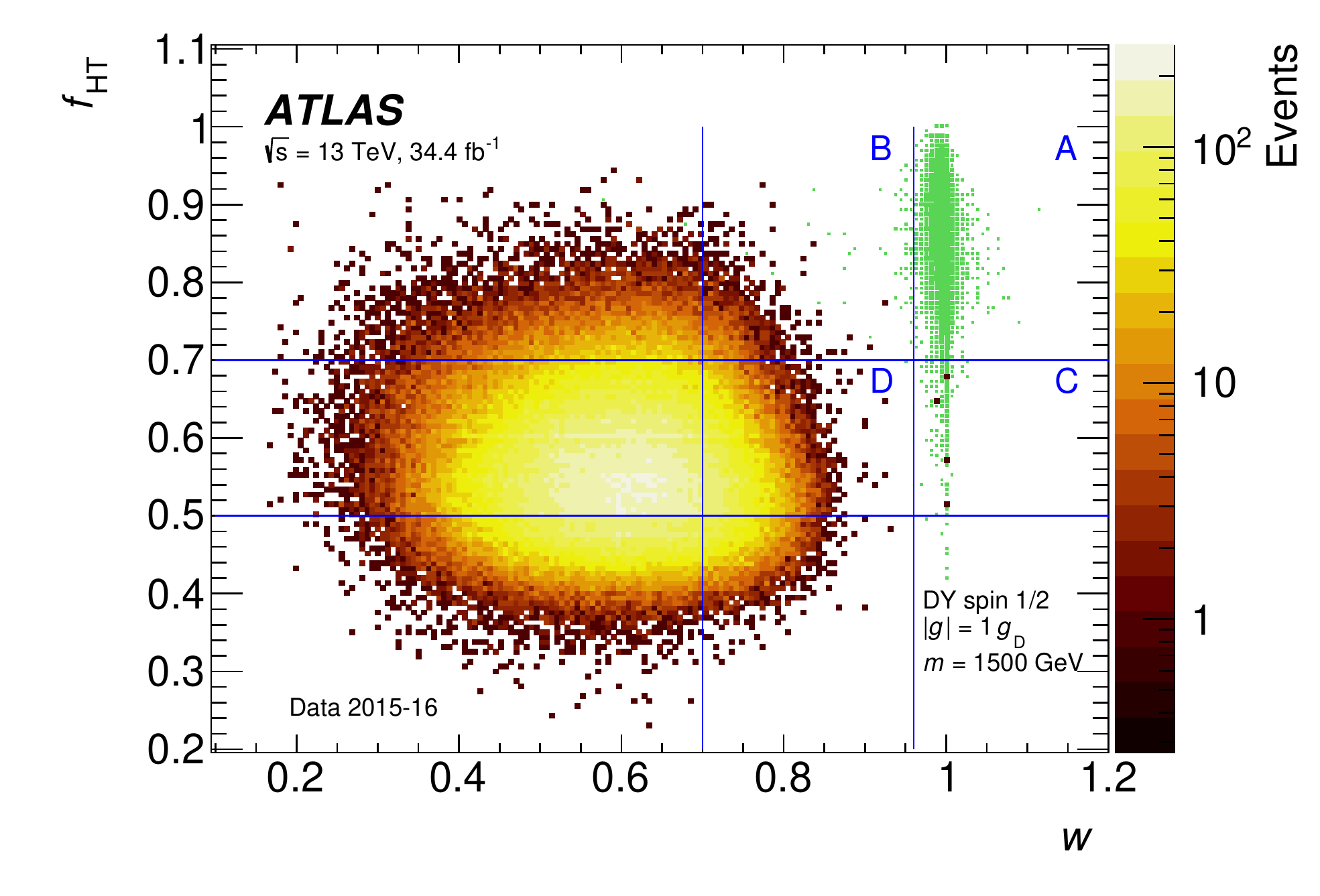} \end{center} \caption{\label{fig:ABCD}
    Two-dimensional distribution of the two discriminating variables
    $f_\text{HT}$ versus $w$ in data (color scale) and a typical HIP
    signal (green squares). The regions for defining the signal (A)
    and for the background estimate and background validation (B, D
    and C) are indicated.}
\end{figure}

Backgrounds are random combinations of rare processes and need to be
estimated directly from the collected data. Examples of background
processes that could yield high $f_\text{HT}$ values include
overlapping charged particles and noise in TRT straws. Processes that
could yield high $w$ values include high-energy electrons and noise in
EM calorimeter cells. The background estimation method relies on the
fact that, in the background near the signal region, $f_\text{HT}$ and
$w$ are largely uncorrelated. Control regions B, C and D are defined
as sidebands in $f_\text{HT}$ and $w$ near the signal region A, as
shown in Figure~\ref{fig:ABCD}. Region~A contains 90\% or more of the
signal for masses below 4000~GeV for all simulated charges except
$|z|=20$, where the fraction is 70\%. The numbers of events observed
in the control regions are $N_\text{B}=1528$, $N_\text{C}=4$ and
$N_\text{D}=30375$, and the expected background is calculated as
$N_\text{A}^\text{exp}=N_\text{B}N_\text{C}/N_\text{D}=0.20\pm0.11 \text{
(stat)} \pm0.40 \text{ (sys)}$. The latter uncertainty accounts for
the fact that $f_\text{HT}$ and $w$ each depend on $\eta$ but in
different ways, resulting in a Pearson correlation coefficient of
10\%. This uncertainty was obtained by binning the $f_\text{HT}$--$w$
plane into $0.025$-unit regions in $\eta$ and determining the maximum
variation of the ratio of the numbers of events in the B and D
regions. Simultaneous fits taking possible signal yields into account
confirm that signal leakage into the B and C control regions cannot
significantly affect the background estimate.  As an additional
cross-check, the B, C and D regions are divided into various
subregions, within which the background estimation is again
performed. The estimated and observed event yields are consistent in
all cases.

\begin{figure*}[tb]
\begin{center}
\includegraphics[width=0.495\linewidth]{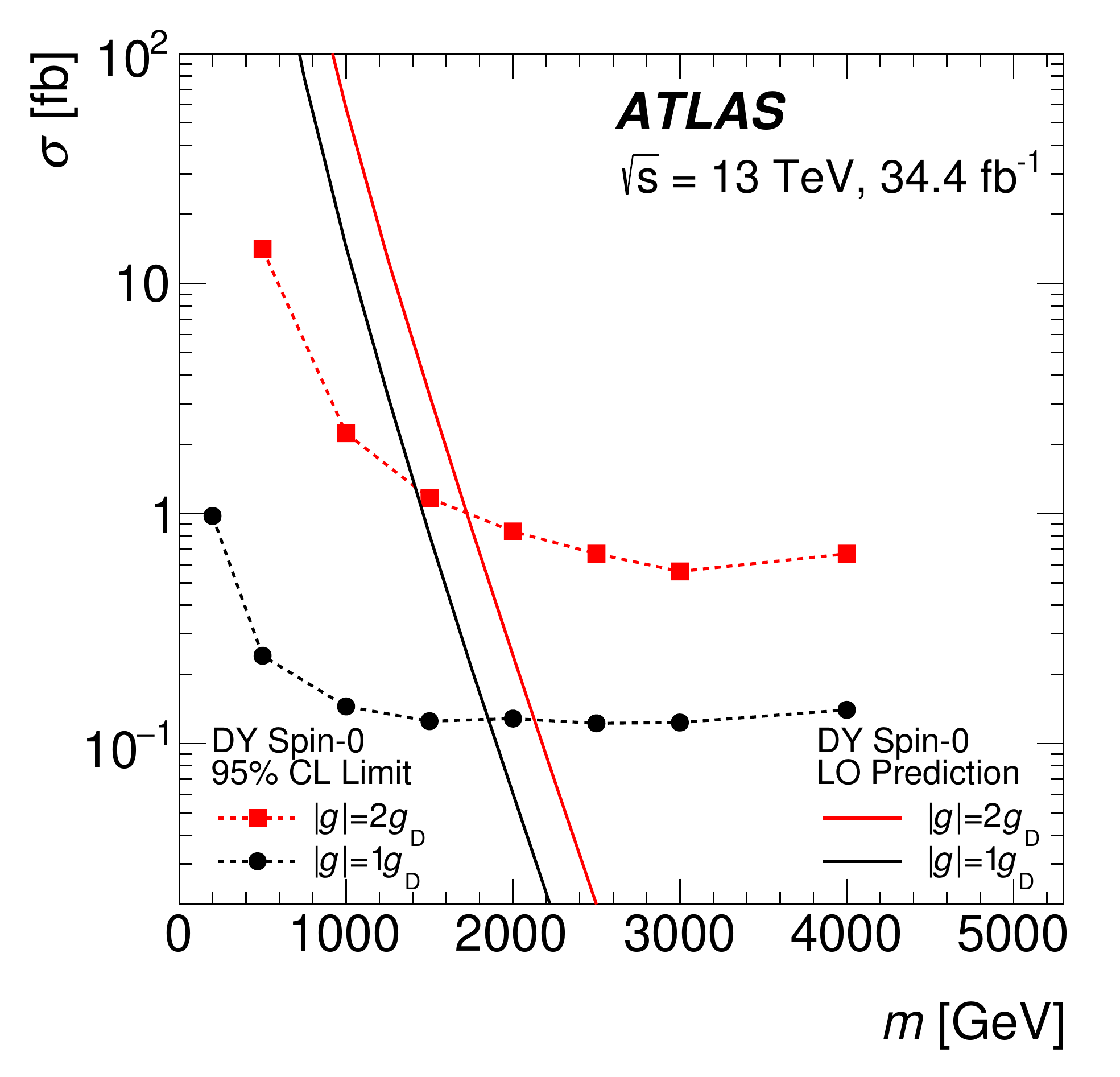}
\includegraphics[width=0.495\linewidth]{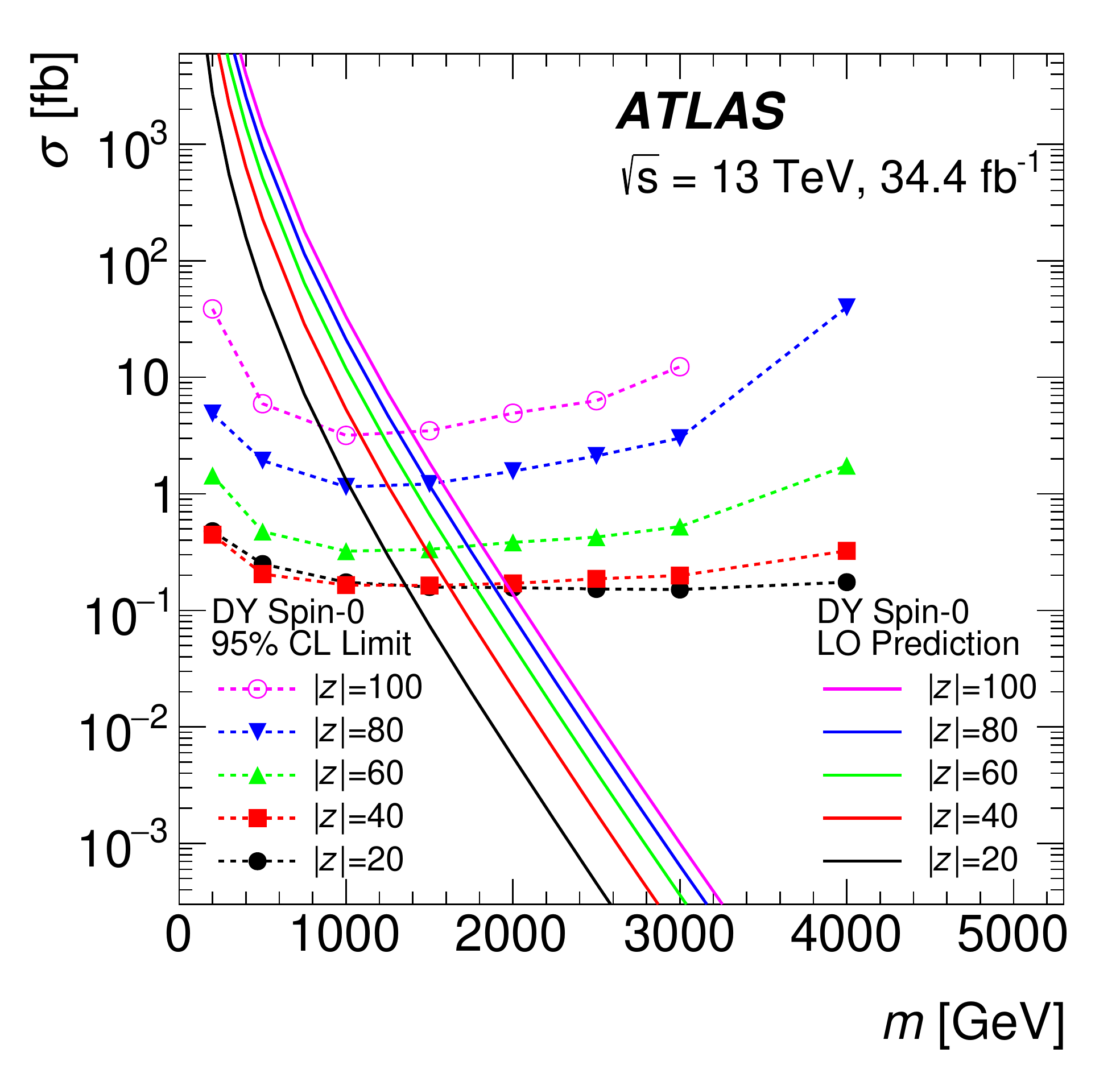}
\includegraphics[width=0.495\linewidth]{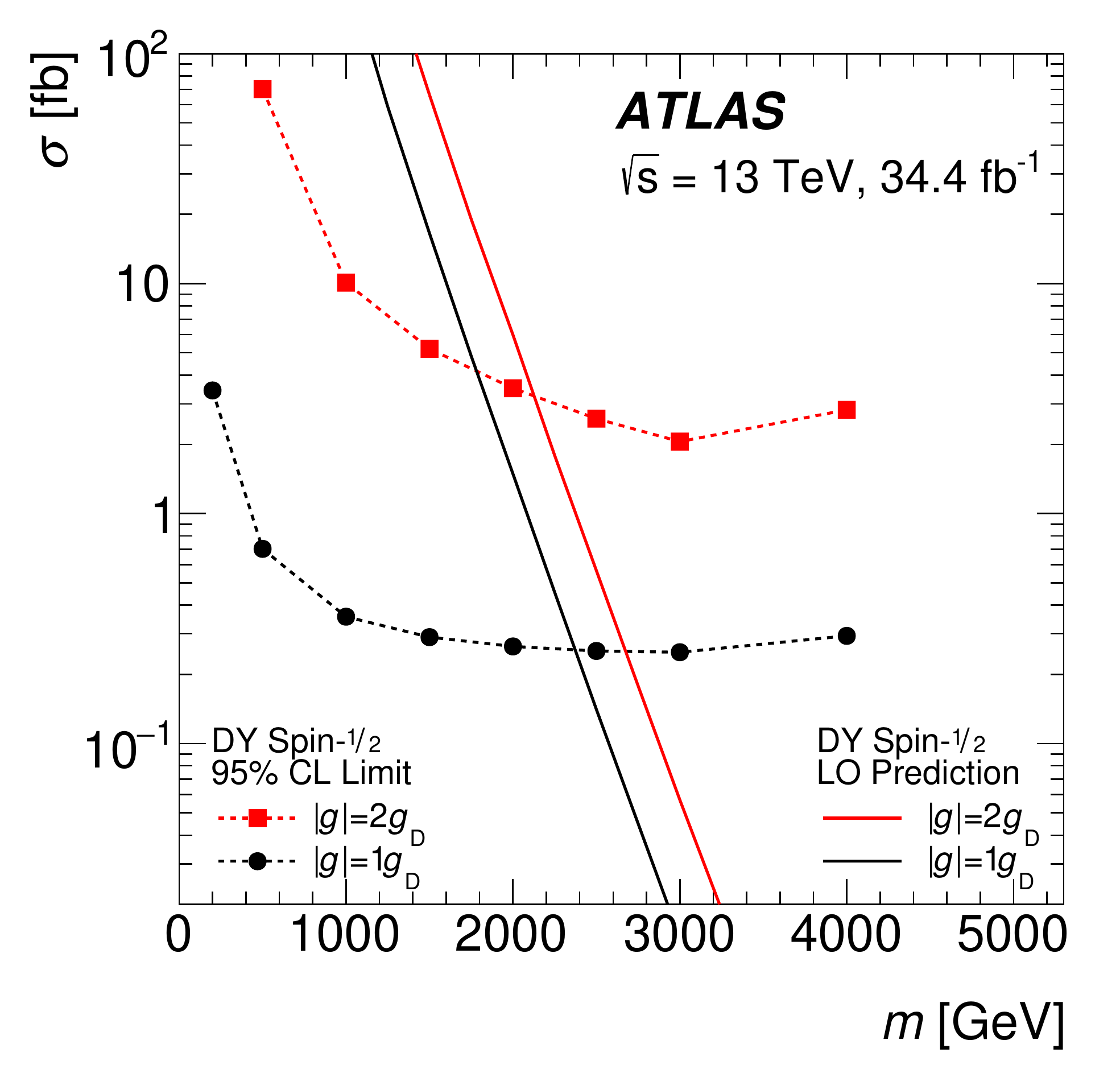}
\includegraphics[width=0.495\linewidth]{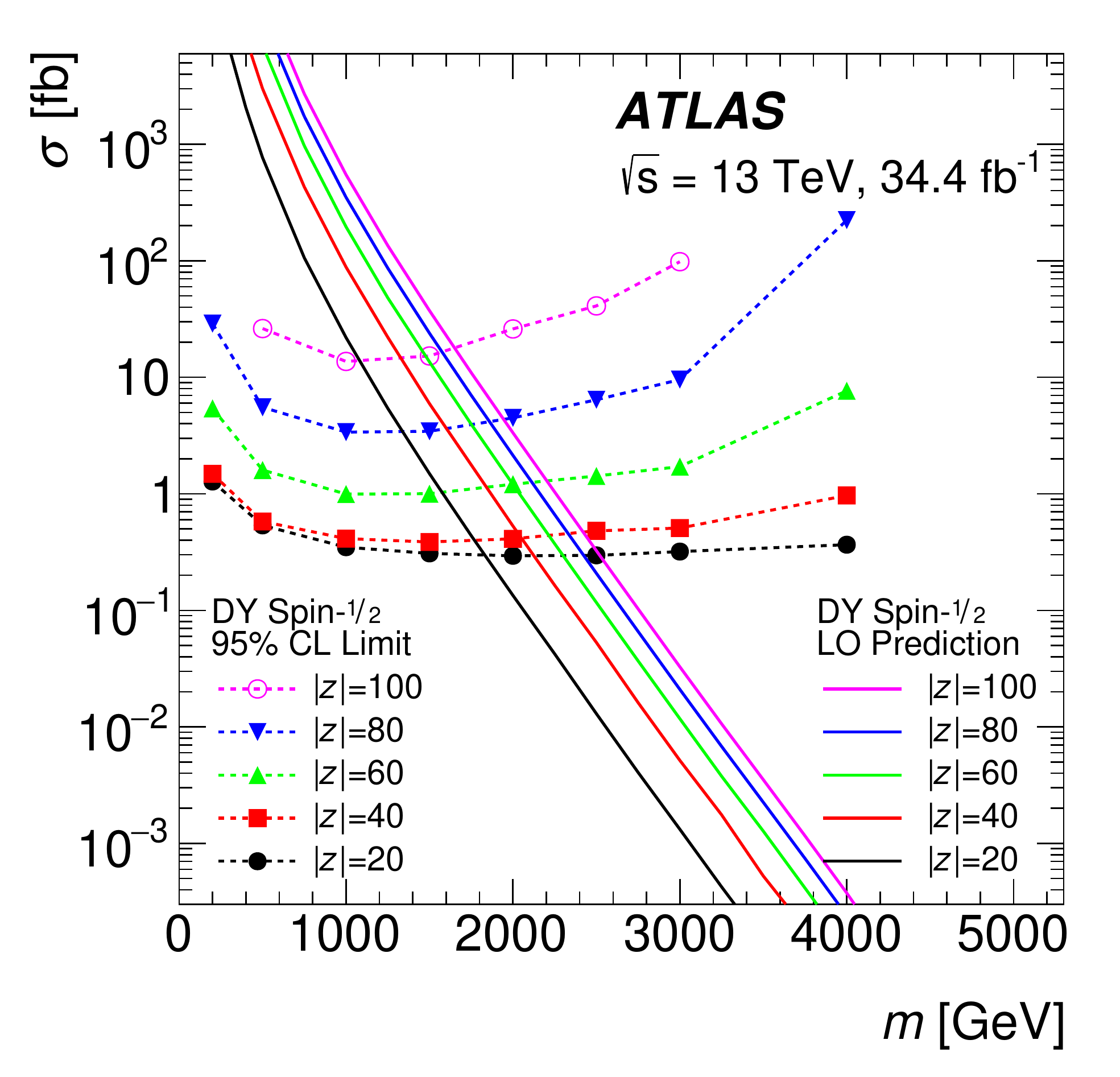}
\caption{\label{fig:limits} Observed 95\% confidence-level upper
    limits on the cross section for Drell--Yan spin-0 (top) and
  spin-1/2 (bottom) monopole (left) and HECO (right) production as a
  function of mass (dashed lines with markers). The limits for spin-0
  HIPs are more stringent than for spin-1⁄2 due to the higher
  selection efficiencies of the former. The theoretical leading-order
  cross sections are overlaid (solid lines).}
\end{center}
\end{figure*}

\begin{table}
\caption{Lower limits on the mass of magnetic monopoles and HECOs (in GeV) at
  95\% confidence level in models of spin-0 and spin-1/2 leading-order DY HIP pair production. }
\label{tbl:masslimits}
\begin{center}
\begin{tabular}{lccccccc}
& \multicolumn{7}{c}{Lower limits on the mass of Drell--Yan magnetic monopoles
and HECOs [GeV]}\\
\hline
\hline
      & $|g|=1g_\text{D}$ & $|g|=2g_\text{D}$ &
      $|z|=20$ & $|z|=40$ & $|z|=60$ & $|z|=80$ & $|z|=100$ \\
\hline
Spin-0   & 1850  & 1725     & 1355     & 1615      & 1625    & 1495   & 1390 \\
Spin-1/2 & 2370  & 2125     & 1830     & 2050      & 2000    & 1860   & 1650 \\
\hline
\hline
\end{tabular}
\end{center}
\end{table}

\FloatBarrier

No event was observed in the signal region A in 34.4~fb$^{-1}$ of
13~TeV $pp$ collision data, consistent with the background
expectation. Thus, 95\% confidence-level (CL) upper limits can be set
on production cross sections for various signal hypotheses, using
estimates of efficiencies and their corresponding uncertainties for
each HIP charge, mass and spin, as well as the uncertainty in the
integrated luminosity (2.2\%, estimated following the methods
discussed in Ref.~\cite{DAPR-2013-01}).  A
$\text{CL}_\text{s}$~\cite{Read:CLs} frequentist framework implemented
in RooStats~\cite{RooStats} is used for hypothesis testing and to
calculate confidence intervals. The resulting limits are shown as a
function of HIP mass in Figure~\ref{fig:limits} for monopoles and
HECOs in the charge ranges where the search is sensitive for DY
production with different spins. The mass dependence of the
cross-section limits arises from a variation of the efficiencies with the HIP
kinetic energy.  The cross-section limits are
relatively insensitive to the systematic uncertainties, which
introduce variations no larger than 12\% across all mass--charge--spin
points. Model cross-section predictions are shown in
Figure~\ref{fig:limits} as solid lines.  
The corresponding mass limits are shown in Table~\ref{tbl:masslimits}.
Given the uncertainty in the predicted cross sections, these mass
limits primarily serve as benchmarks for comparison with other
experiments. The MoEDAL experiment~\cite{MoEDAL2019} is able to set
slightly stronger monopole mass limits because they consider
photon-fusion pair production in addition to the Drell--Yan
mechanism. For most masses, the present cross-section limits obtained
for magnetic charge $|g|=2g_\text{D}$ surpass by one to two orders of
magnitude the best previous constraints, also set by
MoEDAL~\cite{MoEDAL2019}. The cross-section limits obtained for HECOs
and for monopoles with $|g|=g_\textrm{D}$ surpass by approximately a
factor of five the best constraints, set by the previous ATLAS
analysis~\cite{EXOT-2014-16}, and access the range $60<|z|\le100$ for
the first time.

 \FloatBarrier



We thank CERN for the very successful operation of the LHC, as well as the
support staff from our institutions without whom ATLAS could not be
operated efficiently.

We acknowledge the support of ANPCyT, Argentina; YerPhI, Armenia; ARC, Australia; FWF, BMWFW, Austria; ANAS, Azerbaijan; SSTC, Belarus; CNPq, FAPESP, Brazil; NSERC, CFI, NRC, Canada; CERN, CERN; CONICYT, Chile; CAS, NSFC, MOST, China; COLCIENCIAS, Colombia; VSC CR, MSMT CR, MPO CR, Czech Republic; DNSRC, DNRF, Denmark; IN2P3-CNRS, CEA-DRF/IRFU, France; SRNSFG, Georgia; MPG, HGF, BMBF, Germany; GSRT, Greece; RGC, Hong Kong SAR, Hong Kong China; Benoziyo Center, ISF, Israel; INFN, Italy; JSPS, MEXT, Japan; JINR, JINR; CNRST, Morocco; NWO, Netherlands; RCN, Norway; MNiSW, NCN, Poland; FCT, Portugal; MNE/IFA, Romania; NRC KI, MES of Russia, Russia Federation; MESTD, Serbia; MSSR, Slovakia; ARRS, MIZ\v{S}, Slovenia; DST/NRF, South Africa; MINECO, Spain; SRC, Wallenberg Foundation, Sweden; Cantons of Bern and Geneva , SNSF, SERI, Switzerland; MOST, Taiwan; TAEK, Turkey; STFC, United Kingdom; DOE, NSF, United states of America. In addition, individual groups and members have received support from CRC, Compute Canada, Canarie, BCKDF, Canada; Marie Sk{\l}odowska-Curie, COST, ERDF, ERC, Horizon 2020, European Union; ANR, Investissements d'Avenir Labex and Idex, France; AvH, DFG, Germany; Herakleitos, Thales and Aristeia programmes co-financed by EU-ESF and the Greek NSRF, Greece; BSF-NSF, GIF, Israel; PROMETEO Programme Generalitat Valenciana, CERCA Generalitat de Catalunya, Spain; Leverhulme Trust, The Royal Society, United Kingdom.

The crucial computing support from all WLCG partners is acknowledged gratefully, in particular from CERN, the ATLAS Tier-1 facilities at TRIUMF (Canada), NDGF (Denmark, Norway, Sweden), CC-IN2P3 (France), KIT/GridKA (Germany), INFN-CNAF (Italy), NL-T1 (Netherlands), PIC (Spain), ASGC (Taiwan), RAL (UK) and BNL (USA), the Tier-2 facilities worldwide and large non-WLCG resource providers. Major contributors of computing resources are listed in Ref.~\cite{ATL-GEN-PUB-2016-002}.


\clearpage
\printbibliography

\clearpage
 
\begin{flushleft}
{\Large The ATLAS Collaboration}

\bigskip

G.~Aad$^\textrm{\scriptsize 101}$,    
B.~Abbott$^\textrm{\scriptsize 128}$,    
D.C.~Abbott$^\textrm{\scriptsize 102}$,    
O.~Abdinov$^\textrm{\scriptsize 13,*}$,    
A.~Abed~Abud$^\textrm{\scriptsize 70a,70b}$,    
K.~Abeling$^\textrm{\scriptsize 53}$,    
D.K.~Abhayasinghe$^\textrm{\scriptsize 93}$,    
S.H.~Abidi$^\textrm{\scriptsize 167}$,    
O.S.~AbouZeid$^\textrm{\scriptsize 40}$,    
N.L.~Abraham$^\textrm{\scriptsize 156}$
% ...
\bigskip
\\

% ...
$^{13}$Institute of Physics, Azerbaijan Academy of Sciences, Baku; Azerbaijan.\\
% ...
$^{40}$Niels Bohr Institute, University of Copenhagen, Copenhagen; Denmark.\\
% ...
$^{53}$II. Physikalisches Institut, Georg-August-Universit\"{a}t G\"ottingen, G\"ottingen; Germany.\\
% ...
$^{70}$$^{(a)}$INFN Sezione di Pavia;$^{(b)}$Dipartimento di Fisica, Universit\`a di Pavia, Pavia; Italy.\\
% ...
$^{93}$Department of Physics, Royal Holloway University of London, Egham; United Kingdom.\\
% ...
$^{101}$CPPM, Aix-Marseille Universit\'e, CNRS/IN2P3, Marseille; France.\\
$^{102}$Department of Physics, University of Massachusetts, Amherst MA; United States of America.\\
% ...
$^{128}$Homer L. Dodge Department of Physics and Astronomy, University of Oklahoma, Norman OK; United States of America.\\
% ...
$^{156}$Department of Physics and Astronomy, University of Sussex, Brighton; United Kingdom.\\
% ...
$^{167}$Department of Physics, University of Toronto, Toronto ON; Canada.\\
% ...
$^{a}$ Also at Borough of Manhattan Community College, City University of New York, New York NY; United States of America.\\
$^{b}$ Also at Centre for High Performance Computing, CSIR Campus, Rosebank, Cape Town; South Africa.\\
% ...
$^{*}$ Deceased

\end{flushleft}




\end{document}